\documentclass[aps,prb,groupedaddress,showpacs,floatfix,twocolumn]{revtex4}
\usepackage{amsmath,amssymb}
\usepackage{graphicx}
\usepackage{dcolumn}
\usepackage{enumerate}
\usepackage{hyperref}

\newcommand{\x}{\mathrm{x}}
\newcommand{\TMMC}{\mbox{$\mathrm{(CH_3)_4NMnCl_3}$}}
\newcommand{\TMCC}{\mbox{$\mathrm{(CH_3)_4NCdCl_3}$}}
\newcommand{\TMMCCd}{\mbox{$\mathrm{(CH_3)_4NMn}_\x\mathrm{Cd}_{1-\x}\mathrm{Cl_3}$}}
\newcommand{\unit}[1]{\;\mathrm{#1}}
\newcommand{\kB}{k_\mathrm{B}}
\newcommand{\muB}{\mu_\mathrm{B}}

\begin{document}
\bibliographystyle{apsrev}
\title{Magnetization steps in a diluted Heisenberg antiferromagnetic chain:
Theory and experiments on TMMC:Cd}
\author{A. Paduan-Filho}\email{apaduan@if.usp.br}
\author{N. F. Oliveira, Jr.}
\author{V. Bindilatti}
\email{vbindilatti@if.usp.br}
\affiliation{Instituto de F{\'\i}sica, Universidade de S{\~a}o Paulo,\\
 C.P. 66.318, 05315--970 S{\~a}o Paulo-SP, Brazil}
\author{S. Foner}
\affiliation{Francis Bitter Magnet Laboratory, Massachusetts
Institute of Technology, Cambridge, MA 02139}
\author{Y. Shapira}
\email{yshapira@granite.tufts.edu} \affiliation{Department of
Physics and Astronomy, Tufts University, Medford, MA 02155}

\date{\today}
\begin{abstract}
A theory for the equilibrium low-temperature magnetization $M$ of
a diluted Heisenberg antiferromagnetic chain is presented. Only
the nearest-neighbor exchange interaction is included, and the
distribution of the magnetic ions is assumed to be random. Values
of the magnetic fields $B_i$ at the magnetization steps (MST's)
from finite chains with 2 to 5 spins (pairs, triplets, quartets,
and quintets) are given for chains composed of spins $S{=}5/2$.
The magnitudes of these MST's as a function of the fraction, $\x$,
of cations that are magnetic are given for any $S$. An expression
for the apparent saturation value of $M$ is derived. The
magnetization curve, $M$ versus $B$, is calculated using the exact
contributions of finite chains with 1 to 5 spins, and the ``rise
and ramp approximation'' for longer chains. An expression for the
low-temperature saturation magnetic field $B_s(n)$ of a finite
chain with $n$ spins is given. Some non-equilibrium effects that
occur in a rapidly changing $B$, are also considered. Some of
these result from the absence of thermal equilibrium within the
sample itself, whereas others are caused by the absence of thermal
equilibrium between the sample and its environment (e.g.,
liquid-helium bath). Specific non-equilibrium models based on
earlier treatments of the phonon bottleneck, and of spin flips
associated with cross relaxation and with level crossings
(anticrossings), are discussed. Magnetization data on powders of
TMMC diluted with cadmium [i.e.,
{$\mathrm{(CH_3)_4NMn}_\mathrm{x}\mathrm{Cd}_{1-\mathrm{x}}\mathrm{Cl_3}$},
with $0.16\leq\x\leq0.50$] were measured at $0.55\;\mathrm{K}$ in
$18\;\mathrm{T}$ superconducting magnets. The field $B_1$ at the
first MST from pairs is used to determine the NN exchange constant
$J$. This $J/kB$ changes from $-5.9\;\mathrm{K}$ to
$-6.5\;\mathrm{K}$ as $\x$ increases from $0.16$ to $0.50$. The
magnetization curves obtained in the superconducting magnets are
compared with simulations based on the equilibrium theory. A
reasonably good agreement is found. Data for the differential
susceptibility, $dM/dB$, were taken in pulsed magnetic fields
($7.4\;\mathrm{ms}$ duration) up to $50\;\mathrm{T}$. The powder
samples were in direct contact with a $1.5\;\mathrm{K}$
liquid-helium bath. Non-equilibrium effects, which became more
severe as $\x$ decreased, were observed. For $\x{=}0.50$ the
non-equilibrium effects are tentatively interpreted using the
``Inadequate Heat Flow Scenario,'' developed earlier in connection
with the phonon bottleneck problem. The more severe
non-equilibrium effects for $\x{=} 0.16$ and $0.22$ are
tentatively attributed to cross-relaxation, and to crossings (more
accurately, anticrossings) of energy levels, including those of
excited states. For $\x{=} 0.16$  (lowest $\x$), no MST's were
observed above $20\;\mathrm{T}$, which is attributed to a very
slow spin relaxation for pairs, compared to a millisecond. A
definitive interpretation of this and some other non-equilibrium
effects is still lacking.
\end{abstract}

\maketitle

\section{INTRODUCTION\label{sec:intro}}
Spin clusters with predominantly antiferromagnetic (AF) interactions exhibit
steps in the equilibrium magnetization as a function of magnetic field.
These magnetization steps (MST's) arise from energy-level crossings which
change the ground state.
They are observed at very low temperatures when only the ground state
contributes to the magnetization $M$.
In recent years MST's have yielded a wealth of information about AF clusters,
first in diluted magnetic materials and later in molecular magnetism.
An overall review of MST's was published recently.\cite{Shapira02jap}
For recent reviews of the magnetic properties of molecular clusters, including
MST's, see Ref.~\onlinecite{Gatteschi00}.

In a molecular crystal the AF clusters are normally all of one type.
The MST's then give values of exchange constants and anisotropy parameters
for that cluster type.
A diluted magnetic material, on the other hand, contains numerous types of
spin clusters.
Different cluster types give rise to different series of MST's.
In addition to exchange constants and anisotropy parameters, the MST's also
give information concerning the populations of the different cluster types.
The populations are related to the magnitudes of the MST's in the different
series.
The results for the cluster populations can be used to check if the
distribution of the magnetic ions is random.

Most previous studies of MST's in diluted magnetic crystals were on
three-dimensional (3D) materials,\cite{Shapira02jap} although some quantum
wells were also studied.\cite{Crooker99}
The present paper, however, is devoted to MST's from a diluted AF Heisenberg
chain (1D).
The material studied is TMMC [chemical formula: \TMMC] which was diluted by
replacing a large fraction of the Mn atoms by Cd.
Powder samples of \TMMCCd, with $\x$ between $0.16$ and $0.5$, were investigated.
In these materials the non-magnetic $\mathrm{Cd}^{2+}$ ions break the chains
of $\mathrm{Mn}^{2+}$ ions into finite segments.

Pure TMMC is probably the closest approximation to an ideal isotropic
(Heisenberg) linear AF chain.
For reviews of its magnetic properties, with extensive references to original
works, see Refs.~\onlinecite{deJongh74,Steiner76,Carlin86,Khan93}.
The 1D magnetic behavior of this compound is due to the crystallographic
structure.
It contains chains of Mn ions which at room temperature are along the $c$-axis
of the hexagonal structure.
The Mn ions in each chain are linked by Cl ions.
The space between the combined $\mathrm{Mn-Cl_3-Mn}$ chains is occupied by
tetramethylammonium groups.

The strongest magnetic interaction in TMMC is the isotropic exchange between
nearest neighbor (NN) $\mathrm{Mn}^{2+}$ ions in the chain.
The NN exchange constant, obtained from various experiments (e.g.,
Ref.~\onlinecite{Hutchings72}), is $J/\kB \cong-6.6\unit{K}$, where $\kB $ is
the Boltzmann constant.
Other intrachain exchange constants are believed to be much smaller, and are
usually neglected.
The exchange interaction between different chains is orders of magnitude
smaller than the intrachain interaction, but is responsible for the
long-range AF order below the N{\'e}el temperature $T_N {=} 0.84\unit{K}$.
The anisotropy in TMMC is mainly due to the dipole-dipole interaction.
It is of the easy-plane type, and is two orders of magnitude smaller than the
dominant exchange interaction.\cite{Walker72}

At room temperature both TMMC and its Cd analog [\TMCC, known as TMCC] have
isomorphous hexagonal structures (space group $P6_{3/m}$).
Crystallographic phase transitions at lower temperatures result in a lower
symmetry, and in small structural differences between TMMC and
TMCC.\cite{Hutchings72,Peercy73,Braud90,Peral00}
These differences are often assumed to be unimportant, although it is
conceivable that even small changes in the crystal structure have some effect
on the magnetic behavior, especially on spin relaxation at low temperatures.

Previous investigations of \TMMCCd\ included measurements of the
susceptibility and the N{\'e}el temperature as a function of $\x$
(Ref.~\onlinecite{Dupas78}).
These results were interpreted theoretically both by the original authors,
Dupas and Renard, and by Harada \textit{et al.}\cite{Harada80}
Susceptibility measurements on a related diluted linear AF chain (DMMC:Cd)
were obtained and interpreted by Schouten \textit{et al.}\cite{Schouten82}
The authors of both Refs.~\onlinecite{Dupas78} and \onlinecite{Schouten82}
noted the difficulties of preparing alloys with uniform Cd concentrations.

\section{EQUILIBRIUM THEORY\label{sec:equilib}}
In this section a theory for the low temperature magnetization $M$ of a
diluted AF Heisenberg chain is presented.
It is assumed that the spin system is in thermal equilibrium with a
constant-temperature heat reservoir.
This equilibrium theory is suitable for interpreting the data that were
obtained in the slowly-varying magnetic fields (``dc fields'') of the
superconducting magnets.
The additional considerations needed to interpret the results obtained in
pulsed fields, of several ms duration, will be discussed in
Sec.~\ref{sec:noneq}

\subsection{The Model\label{subsec:model}}
MST's from finite AF chains were predicted decades
ago,\cite{Bonner64,Parkinson85} but some more recent theoretical results are
also useful.
The simplest model for MST's in a diluted magnetic material is the single--$J$
cluster model.\cite{Shapira02jap}
It includes only the largest isotropic exchange constant $J$ and the Zeeman
energy.
Other exchange constants, and all anisotropies, are ignored.
This model, with the NN intrachain exchange constant chosen as $J$,
is expected to be a good starting point for \TMMCCd\ (hereafter, TMMC:Cd).
All cluster models  are applicable only when $\x$ is not too high.
However, for  a diluted magnetic chain (1D) the single--$J$ model it is
expected to hold at least up to $\x{=}0.5$.
All the samples in the present study are in this range.

In the single--$J$ model the magnetic clusters are finite chains, each
consisting of $n$ coupled spins. These clusters are treated as independent.
The total magnetization $M$ is then the sum of the magnetizations of finite
chains with different $n$.
Let $\mu_n$ be the average magnetic moment of a chain with $n$ spins, and let
$N_n$ be the number of ``realizations,'' per kg, of a chain with $n$ spins.
($N_n$ may also be called the ``population,'' per kg, of finite chains with $n$
spins).
The magnetization per kg is then
\begin{equation}\label{eq:one}
                M =  \sum N_n\mu_n.
\end{equation}

\begin{figure}[tb]\begin{center}
\includegraphics[width=75mm, keepaspectratio=true, clip]{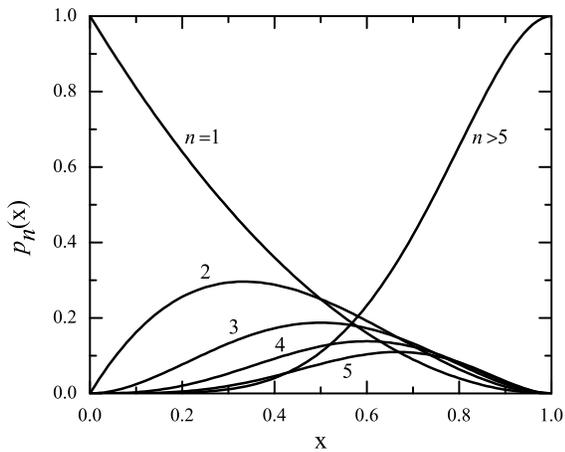}\\
\end{center}\caption{\label{fig1}
The probability $P_n$ that a magnetic ion is in a finite linear chain with
$n$ spins.
Results for $n\leq5$ are plotted as a function of the fraction $\x$ of cations
that are magnetic.
Also plotted is the probability $P_{>5}$ that a magnetic ion is in a cluster
with more than 5 spins.
}
\end{figure}

If $P_n$ is the probability that a spin is in a chain with $n$ spins, and if
$N_\mathrm{total}$ is the total number of spins per kg, then
\begin{equation}\label{eq:two}
                N_n = N_\mathrm{total} P_n /n.
\end{equation}
The probabilities $P_n$ are obtained from well known results,\cite{note18}
\begin{equation}\label{eq:three}
                P_n = n\x^{n-1}(1-\x)^2.
\end{equation}
This result assumes a random distribution of the magnetic ions.
This crucial assumption is discussed later. Figure~\ref{fig1} shows the
probabilities $P_n$ for $n\leq5$, and the probability $P_{>5}$ that a spin is
in a finite chain $n>5$, i.e.,
\begin{equation}\label{eq:four}
                P_{>5} =  1 -\sum_1^5 P_n = (6-5\x)\x^5.
\end{equation}

From Eqs.~(\ref{eq:one}--\ref{eq:three}), the magnetization $M$ is given by
the infinite sum
\begin{equation}\label{eq:five}
                M =  N_\mathrm{total}\sum_n \x^{n-1}(1-\x)^2 \mu_n.
\end{equation}

In practice, exact results for the average magnetic moment $\mu_n$ are
available only for values of $n$ which are not too large (short chains).
The infinite sum in Eq.~(\ref{eq:five}) is therefore truncated at some maximum
value of $n$, called $n_\mathrm{max}$.
Chains with $n>n_\mathrm{max}$ are treated using an
approximation.\cite{Shapira02jap}
In the present work we chose $n_\mathrm{max} {=} 5$ because exact results for
chains with up to five $\mathrm{Mn}^{2+}$ ions ($S {=} 5/2$) were readily
available from previous works.\cite{Shapira02jap,Gratens01prb}

When $\x$ is below $0.50$, less than $11\%$ of the spins are in chains with $n>5$.
For this range of $\x$, the ``rise and ramp'' approximation\cite{Shapira02jap}
can be used for the {\em total} contribution from chains with $n>5$ [i.e., it
approximates the remainder after the sum in Eq.~(\ref{eq:five}) is truncated].
The approximation smoothes the $B$-dependence of this remainder, i.e., MST's
from chains with $n>5$ are not resolved. This approach (exact treatment for
$n\leq5$, and an approximation for $n>5$) proved to be adequate for
interpreting the data obtained in dc magnetic fields.
However, the interpretations of some data obtained in pulsed magnetic fields
also used available theoretical results\cite{Lou02} for $n>5$.

To implement the rise and ramp approximation, the so-called
``short chain model'' (Ref.~\onlinecite{Schouten82}) was adopted.
This model is appropriate for $\x<0.5$ because more than $99.4\%$ of the spins
are in chains with $n\leq10$, and more than $99.9\%$ are in chains with
$n\leq14$.
A chain with 14 spins is still short enough to be described by that model.
In the short-chain model the ground state at $B {=} 0$ has total spin
$S_T(0){=}0$ when $n$ is even, and $S_T(0){=}S$  if $n$ is odd.
This simple result does not hold for 3D materials, which is one of the reasons
why the rise and ramp approximation is simpler and much more accurate for
diluted chains than for diluted  3D materials.

\subsection{\label{subsec:shape}
Qualitative shape of the magnetization curve at low temperatures}
\subsubsection{Clusters of one type}
Consider first an ensemble of identical finite chains, all with the same $n$.
(In the language of Ref.~\onlinecite{Shapira02jap}, this is an ensemble of
``realizations'' of a chain ``type'' with $n$ spins.)
At low temperatures, $k_BT\ll|J|$, the qualitative variation of $\mu_n$ with
$B$ depends on whether $n$ is even or odd. When $n$ is odd the
zero-field-ground-state total spin $S_T(0){=}S$ aligns rapidly at low $B$.
This alignment is given by  the Brillouin function (BF) for spin $S$.
The rapid rise of $\mu_n$ at low $B$ ends when the BF approaches saturation.
After the BF saturates, there is a magnetic-field interval in which $\mu_n$
is nearly constant.
At still higher $B$ a series of MST's appears.
The fields at these MST's depend on $n$.
Once this series of MST's is completed, $\mu_n$ reaches its true saturation
value $\mu_{n,\mathrm{max}}{=} ng\muB  S$.

When $n$ is even, $S_T(0) {=} 0$.
Therefore, no initial fast rise of $\mu_n$ occurs at low $B$, in contrast to
the case of odd $n$.
However, at high magnetic fields a series of MST's still appears.
At the completion of this series, $\mu_n$  reaches its true saturation value.

\subsubsection{Total magnetization $M$}
Chains with all values of $n$ contribute to the total magnetization $M$.
When $k_BT\ll|J|$,  the chains with odd values of $n$ produce a fast rise of
$M$ at low $B$.
This rise follows the BF for spin $S$. After this fast rise is completed, and
before the appearance of the first MST of significant size, there exists a
field interval in which $M$ stays approximately constant.
This (nearly) constant value, $M_s$, is the ``apparent saturation value''
(see Ref.~\onlinecite{Shapira02jap}).
At still higher fields, MST's series from chains with $n\geq2$ appear.
Once the last MST of significant size is completed, $M$ reaches its true
saturation value $M_0$.
As discussed later, the magnetic field required to saturate the magnetization
of a chain remains finite even when $n\to\infty$, so that $M$ reaches true
saturation at a finite $B$.

\subsection{Apparent saturation value}
The true saturation value of the magnetization is
\begin{equation}\label{eq:six}
   M_0 = N_\mathrm{total}g\muB  S.
\end{equation}
The apparent saturation value is
\begin{equation}\label{eq:seven}
                M_s =\sum_{n=\mathrm{odd}}  N_\mathrm{total} \x^{n-1}(1-\x)^2
g\muB  S,
\end{equation}
where the sum is only over odd $n$.
Therefore,
\begin{equation}\label{eq:eight}
                    M_s/M_0 =\sum_{n=\mathrm{odd}}  \x^{n-1}(1-\x)^2.
\end{equation}
This infinite geometric series can be summed,\cite{note21}
\begin{equation}\label{eq:nine}
                    M_s/M_0 = (1-\x)/(1+\x).
\end{equation}

It is noteworthy that an exact analytical expression for $M_s$ was not
obtained for 3D materials.\cite{Shapira02jap}
A rough approximation was then used for the net contribution from clusters
with $n>n_\mathrm{max}$.
This approximation was used only for $\x\lesssim0.1$.
In contrast, Eq.~(\ref{eq:nine}) for a diluted chain is exact, within the
framework
of the short-chain model. It should be very accurate for $\x\le 0.5$.

\subsection{MST's from Chains with $n\le5$\label{subsec:magclusters}}
All chains with $n\ge2$ give rise to MST's.
The magnetic fields $B_i$ at the MST's from chains composed of spins $S{=}5/2$
were given earlier for $n{=}2$ (pairs, or dimers), $n{=}3$ (triplets, or
trimers), $n{=}4$ (quartets, or tetramers), and $n{=}5$ (quintets, or
pentamers).\cite{Gratens01prb}
For completeness, the values of the reduced fields $b_i = g\muB  B_i/|J|$ are
repeated here.

For  $n{=}2$ there are five MST's at $b_i {=} 2, 4, 6, 8, 10$.
For $n{=}3$ there are five MST's are at $b_i {=} 7,$ $9$, $11$, $13$, $15$.
For  $n{=}4$  there are ten MST's at $b_i {=} 0.95$, $2.04$, $3.39$, $5.02$,
$6.87$, $8.85$, $10.88$, $12.94$, $15.00$,
$17.07$. For  $n{=}5$ there are ten MST's at $b_i {=} 4.62$, $5.89$, $7.18$,
$8.49$, $9.86$, $11.29$, $12.83$, $14.48$, $16.24$, $18.09$.

\begin{figure}[tb]\begin{center}
\includegraphics[width=75mm, keepaspectratio=true, clip]{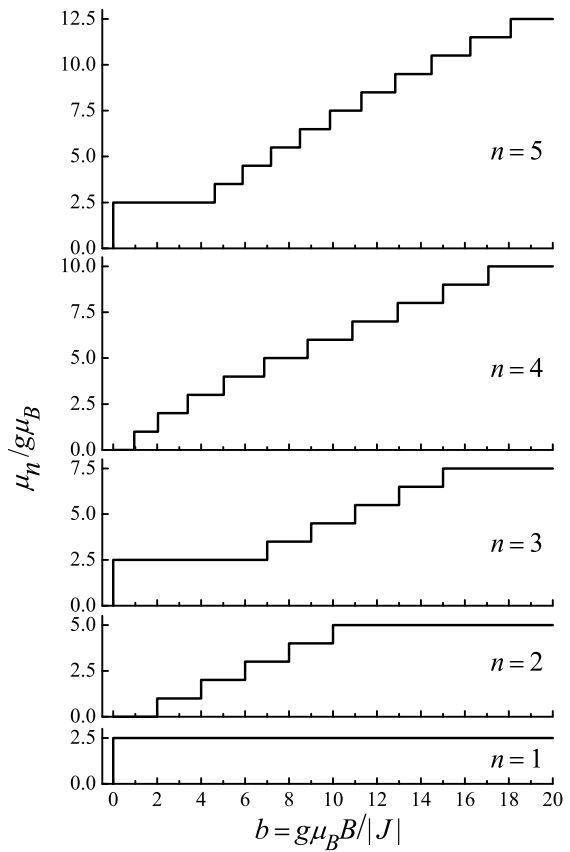}\\
\end{center}\caption{\label{fig2}
Magnetic moment per cluster, $\mu_n$, for chains with $n\le5 $ at $T{=}0$,
plotted  as a function of the reduced field $b{=}g\muB  B/|J|$.
}
\end{figure}

Figure~\ref{fig2} shows the zero-temperature values of $\mu_n$ as a
function of $b$ for $n {=} 1, 2, 3, 4, 5$.
At finite $T$ the ground state of a chain is not the only contributor to
$\mu_n$.
It is then necessary to include all energy levels in the calculations of
$\mu_n$.
The procedures for such calculations were discussed in
Ref.~\onlinecite{Shapira02jap}.

\subsection{\label{subsec:rise}
Rise and ramp approximation for chains with $n>5$}
The total contribution of chains with $n>5$ to $M$ is approximated
by a sum of two terms:
1) a fast rise at low $B$,  which follows the BF for spin $S$, and
2) a linear ``ramp'' from $B {=} 0$ up to an effective saturation field $B_s$.

The fast rise at low $B$ is due to chains with odd $n$, starting with $n{=}7$.
Its magnitude is
\begin{equation}\label{eq:ten}
            (\Delta M)_\mathrm{rise} = M_0 \x^6(1-\x)/(1+\x).
\end{equation}
The linear ramp approximates the superposition of numerous MST's from all
chains with $n>5$.
The ramp is given by
\begin{equation}\label{eq:eleven}
            M_\mathrm{ramp} =
            \left\{
            \begin{array}{l}
                (\Delta M)_\mathrm{ramp} (B/B_s),\textrm{\ for\ }B\le B_s\\
                (\Delta M)_\mathrm{ramp},\quad \textrm{\ for\ } B > B_s,
            \end{array}
            \right.
\end{equation}
where
\begin{eqnarray}\label{eq:twelve}
            (\Delta M)_\mathrm{ramp} &=& P_{>5} M_0 - (\Delta
M)_\mathrm{rise}\nonumber\\
            &=&M_0\x^5(6-4\x^2)/(1+\x).
\end{eqnarray}

The reduced field $b_s(n)$ where the magnetization of a finite chain with $n$
spins reaches saturation at $T{=}0$ increases with $n$.
However, in the limit $n\to\infty$ it is still
finite,\cite{Parkinson85,Gratens01prb} namely, $b_s(n{=}\infty) = 8S$.
It can be shown that the $n$-dependence of $b_s(n)$ is given by the equation
\begin{equation}\label{eq:thirteen}
                      b_s(n) =  8S \cos^2(\pi/2n). 
\end{equation}
For chains with $n>5$ the change of this function is only $7\%$.
In the present case of $S{=}5/2$, $b_s(n)$ changes from $18.7$ to $20.0$ when
$n$ increases from  $6$ to $\infty$.
The value $b_s {=} 19$,  corresponding to
\begin{equation}\label{eq:fourteen}
               g\muB  B_s  =  19 |J| ,
\end{equation}
will be used in Eqs.~(\ref{eq:eleven}) for the ramp.

\begin{figure}[tb]\begin{center}
\includegraphics[width=75mm, keepaspectratio=true, clip]{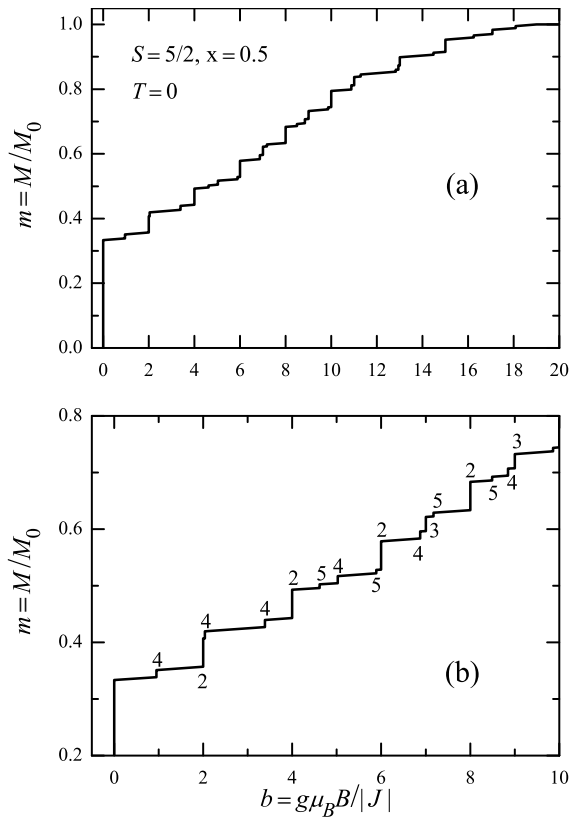}\\
\end{center}\caption{\label{fig3}
a) Calculated zero-temperature equilibrium magnetization of a
diluted linear chain with $\x{=}0.5$ as a function of the reduced
field $b$. These results are for $S{=}5/2$. b) Expanded view of
part (a) for the range of reduced fields relevant to the present
work. The values of $n$ for the chains responsible for each MST
are indicated. }
\end{figure}

Figure~\ref{fig3}(a) shows the predicted zero-temperature magnetization
curve for $S {=} 5/2$ when $\x{=}0.50$.
Figure~\ref{fig3}(b) is an expanded view for the range of magnetic fields
relevant to the present experiments.
The integers in this figure are the values of $n$ for the finite chains
responsible for each MST.

\subsection{\label{subsec:random}Random Distribution}
Some of the preceding results used the probabilities $P_n$ for a random
distribution of the magnetic ions.
Although a random distribution is often found in diluted magnetic materials,
non-random distributions were also observed.\cite{Shapira02jap}
In the case of TMMC:Cd, difficulties of obtaining a uniform Mn distribution
were reported in the literature.\cite{Dupas78,Schouten82}
Therefore, the possibility of a non-random distribution cannot be ignored.
The effects on the magnetization curve caused by departures from a random
distribution were discussed in Ref.~\onlinecite{Shapira02jap}.
For example, a tendency of the magnetic ions to cluster together is expected
to decrease $M_s$.

\section{NON-EQUILIBRIUM  EFFECTS\label{sec:noneq}}
\subsection{Non-Equilibrium Effects\label{subsec:noneqeffects}}
Experiments are normally conducted with the sample in contact with, or near,
a thermal reservoir of constant temperature, e.g., a liquid helium bath at a
temperature $T_\mathrm{bath}$.
In some experiments the sample is not in thermal equilibrium either within
itself and/or with the thermal reservoir.
That is, the time for reaching complete equilibrium (both internal and with
the reservoir) is not short compared to the time of the experiment.
Such non-equilibrium cases require special considerations.

Thermal equilibrium is usually maintained if 1) the magnetic field $B$ is swept
slowly (``dc fields''), and 2) the sample is in good thermal contact with the
reservoir.
These conditions are often fulfilled in superconducting magnets when the
sample is immersed in liquid helium.
There are, however, exceptional cases of non-equilibrium behavior even for
slowly varying $B$ and good thermal contact.
These have been discussed extensively in connection with macroscopic quantum
tunnelling.
A well known example is $\mathrm{Mn_{12}}$-acetate.%
\cite{Gatteschi00,Friedman96,Thomas96,Chudnovsky98,Barbara99}

Departures from thermal equilibrium with the reservoir due to imperfect
thermal contact were observed and discussed for a wide range of sweep rates,
$dB/dt$, from  typical sweep rates in ``dc magnets'' to the very fast rates in
pulsed magnets.\cite{Shapira89,Bindi91,Foner94,Shapira99,Chiorescu00,Nakano01,%
Shapira01,Waldmann02,Inagaki03}
The extreme case of a sample isolated from the thermal reservoir (adiabatic
conditions) was discussed by Wolf long ago, assuming thermal equilibrium
within the sample.\cite{Wolf59}
The magnetocaloric effect leads to cooling when any one of the MST's is
approached.
Such ``cooling by adiabatic magnetization'' has been observed many times in
both dc fields and pulsed fields.\cite{Shapira89,Shapira99,Shapira01,%
Waldmann02,note36}

\subsection{\label{subsec:neqmodels}
Some non-equilibrium models for pulsed magnetic fields}
\subsubsection{\label{subsubsec:class}
Classification of non-equilibrium behaviors}
Non-equilibrium behavior in pulsed fields of milliseconds duration is relevant
to the interpretation of the present pulsed field data.
For this purpose it will be useful to distinguish between three types of
non-equilibrium situations:
\begin{enumerate}
\item 
The spin-lattice relaxation is fast enough so that thermal equilibrium within
the sample is established in a time which is very short compared to the pulse
duration.
The non-equilibrium behavior is then due to an inadequate heat flow between
the liquid-helium bath and the sample, i.e., the sample-bath equilibration
time is not short compared to the pulse duration.
The ``Inadequate Heat Flow'' models are appropriate for this situation.
\item 
The time for establishing thermal equilibrium within the sample is not short
compared to the pulse duration, but there is adequate sample-to-bath heat flow.
The non-equilibrium is then governed by the slow spin-lattice relaxation
processes. There are many such processes.
Here, only spin-flips associated with level crossings or with cross-relaxation
(CR) will be discussed.\cite{Ajiro98,Wernsdorfer02}
\item 
Both the time for reaching equilibrium within the sample, and the time for
reaching a sample-to-bath equilibrium are not short compared to the pulse
duration. Some such cases were considered in the
literature.\cite{Chiorescu00,Nakano01,Waldmann02,Giraud01}
\end{enumerate}

\subsubsection{Inadequate Heat Flow (IHF) Models\label{subsubsec:IHF}}
The phonon bottleneck (PB) phenomenon has been known for a
long time.\cite{vanVleck41,Abragam70}
Treatments of this phenomenon in the context of the magnetization process of
magnetic clusters led to the development of several IHF models.%
\cite{Bindi91,Shapira99,Chiorescu00,Nakano01,Waldmann02,Inagaki03}
The common feature of these models is that the spin and phonon subsystems,
within the sample, are very nearly in thermal equilibrium.
The spin temperature $T_s$, the phonon temperature, and the sample temperature
are the same.
However, the sample and the helium bath are not in thermal equilibrium
($T_s\neq T_\mathrm{bath}$).
Different IHF models treat the sample-to-bath heat flow differently.
For a solid sample in contact with a liquid-helium bath the heat flow was
assumed to be limited by the Kapitza resistance.%
\cite{Bindi91,Shapira99,Inagaki03}
Many of the qualitative features of the magnetization curve are common to all
IHF models; they do not depend on the detailed treatment of the heat flow.

Because the sample is internally in equilibrium, the equilibrium theory in
Sec.~\ref{sec:equilib} still applies.
The absence of equilibrium with the helium bath enters only in the time
dependence (and, hence, the $B$-dependence) of the spin temperature $T_s$.
This dependence has been calculated using various IHF models.%
\cite{Bindi91,Shapira99,Chiorescu00,Nakano01,Waldmann02}
Approaching the energy-level crossing (strictly, anticrossing) associated with
each MST, $T_s$ decreases.
After passing through the anticrossing region, $T_s$ increases.
If some heat flows between the sample and the bath then the latter increase
is large enough that $T_s$ is temporarily above $T_\mathrm{bath}$.
Miyashita and co-workers\cite{Nakano01} have called this behavior of $T_s$
``the magnetic Foehn effect.''

The $B$-dependence $T_s$ can lead to the following qualitative effects:
1) The $dM/dB$ peaks are narrower than the thermal width.
2) The $dM/dB$ peaks are asymmetric, i.e., the rise of $dM/dB$ as the MST is
approached is faster than the fall after passing through the MST.
This is true both for increasing and decreasing $B$.
3) The magnetization and $dM/dB$ exhibit hysteresis.
4) Under some conditions, a small ``satellite'' MST appears after the main MST.
All these effects have been observed experimentally.

\subsubsection{Cross Relaxation (CR) Model\label{subsubsec:cross}}
A severe non-equilibrium behavior, not explainable by IHF models, was observed
in pulsed field experiments by Ajiro \textit{et al.}\cite{Ajiro98}
It was interpreted in terms of CR between pairs and singles, and also between
the pairs themselves. More recently, CR was discussed in the context of
tunneling theory.\cite{Wernsdorfer02,Giraud01}
CR can involve both ground and excited states.

CR is one of the mechanisms of spin relaxation. It involves simultaneous spin
flips in weakly coupled clusters. The model of Ajiro \textit{et al.} also
includes a single spin flip in only one of the clusters.
The later is actually not a CR process, and is better described as tunneling.
In this model the spin relaxation rate is appreciable only at some values
of $B$.

The simplified picture used in Ref.~\onlinecite{Ajiro98} ignored small level
repulsions near level crossings.
(Level repulsion is included in more detailed models that are based on
tunneling.\cite{Wernsdorfer02,Giraud01})
In this simplified picture the relaxation rate can be fast only if
simultaneous CR spin flips, or a single spin flip, do not change the total
energy of the spin system. For a single spin flip, in one cluster, this
happens at energy-level crossings for this cluster.
These level crossings include those of excited states.
For CR between two weakly coupled clusters the energy-level
separation in one cluster should match a level separation in the other.
The two clusters may be of the same type (e.g., two coupled pairs) or of
different types (e.g., a pair and a single).
A similar criterion for energy-level separations applies to CR between three,
or more, coupled clusters.

In the equilibrium theory, the $dM/dB$ peaks associated with MST's
from $\mathrm{Mn}^{2+}$ pairs occur at fields
\begin{equation}\label{eq:fifteen}
                g\muB  B_i  = 2i|J|,
\end{equation}
where  $i {=} 1,2,\dots,(2S{+}1)$. In the model of Ajiro \textit{et al.}
these peaks from pairs are called the ``fundamental'' peaks, and they are
labeled as $\mathrm{P}_i$.
Specializing to pairs composed of $\mathrm{Mn^{2+}}$ ions, there are five
such peaks.
In addition to the fundamental peaks, other peaks are also predicted.
The most pronounced are the ``second harmonic'' peaks $\mathrm{P}_{m/2}$ at
fields $B_m {=} m(B_1/2)$, with $m {=} 1 \text{\ to\ }10$.
The five peaks for odd $m$ were observed clearly by Ajiro \textit{et al.}
The peaks for even $m$ coincide with the fundamental peaks.

In addition to the second-harmonic peaks, fifteen third harmonic peaks
$\mathrm{P}_{k/3}$ at $B_k {=}k(B_1/3)$, were predicted.
Many of these were also observed. Fourth and sixth harmonic peaks were also
discussed.
These results are from a model which involves only singles and pairs.
The model can be extended to include other clusters.

\begin{figure}[tb]\begin{center}
\includegraphics[width=75mm, keepaspectratio=true, clip]{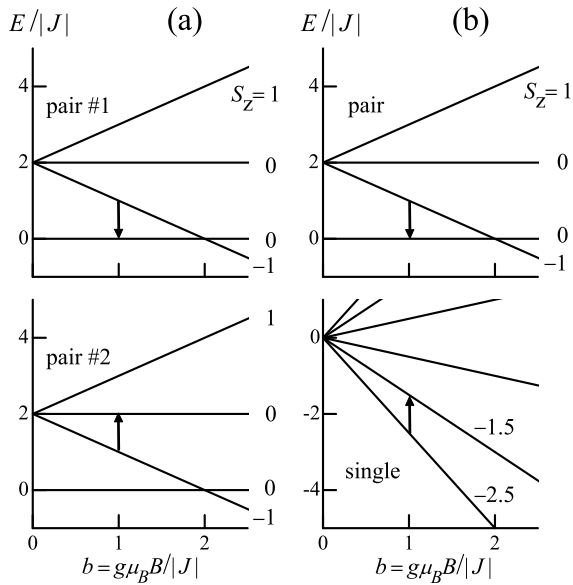}\\
\end{center}\caption{\label{fig4}
Some CR spin-flip transitions which may lead to the $\mathrm{P}_{1/2}$
peak in $dM/dB$.
This peak is at $B{=}|J|/g\muB  $.
a) Simultaneous spin flips in two pairs.
b) Simultaneous spin flips in a pair and in a single.
}
\end{figure}

\begin{figure}[tb]\begin{center}
\includegraphics[width=75mm, keepaspectratio=true, clip]{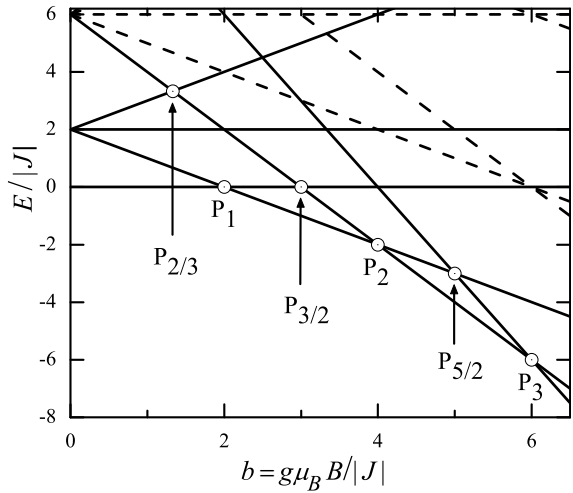}\\
\end{center}\caption{\label{fig5}
Some energy-level crossings which may contribute to the
``fundamental'' peaks $\mathrm{P}_1$, $\mathrm{P}_2$, and $\mathrm{P}_3$, and
to the ``harmonic'' peaks $\mathrm{P}_{3/2}$ , $\mathrm{P}_{5/2}$, and
$\mathrm{P}_{2/3}$.
The $\mathrm{P}_{k/r}$  peak is the  $k$-th peak of the $r$-th harmonic in the
model of Ajiro \textit{et al.}\cite{Ajiro98}
}
\end{figure}

Some of the simultaneous spin-flip transitions which can give rise to
$\mathrm{P}_{1/2}$ (the first of the ten second-harmonic peaks) are shown in
Figs.~\ref{fig4}(a) and \ref{fig4}(b).
Note that in either case, each of the two simultaneous spin flips increase the
component of the spin along $B$ by one unit.
Figure~\ref{fig5} shows a few level crossings that
may contribute to various peaks: the first three fundamental peaks
$\mathrm{P}_1$, $\mathrm{P}_2$, $\mathrm{P}_3$; the third and fifth
second-harmonic peaks, $\mathrm{P}_{3/2}$ and $\mathrm{P}_{5/2}$;
and the second third-harmonic peak, $\mathrm{P}_{2/3}$.
Cross relaxation processes may also contribute to some of the same peaks,
e.g., to $\mathrm{P}_{2/3}$.

\section{EXPERIMENTAL TECHNIQUES\label{sec:experimental}}
\subsection{Sample preparation\label{subsec:sample}}
The procedure of growing crystals of TMMC:Cd followed
Ref.~\onlinecite{Dupas78}.
The samples were grown by evaporation from water solutions of
$\mathrm{MnCl_2.4H_2O}$, $\mathrm{(CH_3)_4NCl}$, $\mathrm{CdCl_2.2H_2O}$,
and some $\mathrm{HCl}$ acid.
The solutions were maintained at $30^\mathrm{o}\mathrm{C}$.
As already noted, the Cd concentration in the crystallized samples is about
$50$ times larger than in the starting solution.\cite{Dupas78,Schouten82}
Because the Cd concentration in the solution decreases rapidly as the crystal
growth progresses, a large solution volume ($300\unit{ml}$) was used to
crystalize a ``product'' of TMMC:Cd with a total mass of about
$100\unit{mg}$.
The large starting volume increased the uniformity of the Cd concentration in
the product.

Physically, the product consisted of many needle-shaped crystals.
The long dimension of a needle (up to $4\unit{mm}$) was parallel to the
hexagonal axis.
The color gradually changed from pink towards white as the Mn concentration $\x$
decreased.
Each of the magnetization measurements, in both dc magnetic fields and in
pulsed fields, used only a portion of the product, typically  $30\unit{mg}$.

The Mn concentration $\x$, was determined from the high-temperature
dc susceptibility, $\chi{=}M/B$, measured using a SQUID
magnetometer. The magnetic field $B$ = $0.1\unit{T}$ was well
within the range where $\chi$ is independent of $B$. In the
temperature range from $150\unit{K}$ to $300\unit{K}$, the data
for $\chi$  were well described by a sum of a Curie-Weiss
susceptibility and a constant due to the diamagnetism of the
lattice. The concentration $\x$ was obtained from the Curie  constant
$C$. Strictly, the Curie-Weiss law is accurate only in the
limit of very high temperatures. The percentage error in the Curie constant,
resulting from the use of data between $150\unit{K}$ and $300\unit{K}$,
depends on the Mn concentration. Based on the results in
Ref.~\onlinecite{Dong88}, for the samples used here the error in $C$,
and hence in $\x$, was less $2\%$. The percentage error in the
Curie-Weiss temperature $\theta$ is larger than for $C$, but
$\theta$ does not enter into the determination of $\x$.

Values of $\x$ for several portions of the $\sim100\unit{mg}$ product obtained
from a single solution were close to each other.
A quantitative comparison between experimental dc magnetization data and
theoretical simulations was carried only for samples $2$, $4$, and $5$ in
Table~\ref{table1}.
To increase the confidence in this comparison, values of $\x$ were determined
for the very same three samples.
As a check, the Mn and Cd weight percents for these three samples were also
determined directly by atomic emission spectroscopy with inductively coupled
plasma (ICP-AES).
The values of $\x$ deduced from ICP-AES were in reasonable agreement with the
values from the susceptibility (see Table~\ref{table1}).

The magnetization of several other samples ($1^*$, $3^*$, $4^*$, and $5^*$ in
Table~\ref{table1}) was also measured in dc fields.
However, for these samples each value of $\x$ is from susceptibility data on a
different portion of the same product.
Samples $4$ and $4^*$ are two different portions of the same product, as are
samples $5$ and $5^*$.

X-ray powder diffraction patterns were obtained at room temperature using
Cu-$\mathrm{K_\alpha}$ radiation. Data were taken on two samples from the
same products as those of samples 1* and 4 (or 4*).
The diffraction patterns for both samples were very similar to the pattern
obtained, with the same  equipment, on pure TMCC ($\x{=}0$).
No additional, or missing, diffraction peaks were observed.
These results are consistent with a single crystallographic phase.

\begin{table}
\caption{\label{table1}
Properties  of the various samples. The Mn concentration $\x$, as determined from
the magnetic susceptibility (Suscept.) and from atomic emission spectroscopy
with inductively coupled plasma (ICP-AES).
The magnetic field $B_1$ is at the first MST from pairs, and the magnetic
field $B_{1\mathrm{QUART}}$ is at the first MST from quartets.
Both fields were determined from dc magnetization data.
The NN exchange constant $J$ was obtained from $B_1$.
}
\begin{ruledtabular}
\begin{tabular}{lddddd}
Sample & \multicolumn{1}{c}{$\x$} & \multicolumn{1}{c}{$\x$}
       & \multicolumn{1}{c}{$B_1$} & \multicolumn{1}{c}{$J/\kB $}
       &\multicolumn{1}{c}{$B_{1\mathrm{QUART}}$}\\
No.    &\multicolumn{1}{c}{\mbox{Suscept.}}
       & \multicolumn{1}{c}{\mbox{ICP-AES}}
       &\multicolumn{1}{c}{\mbox{(T)}}
       &\multicolumn{1}{c}{\mbox{(K)}}&\multicolumn{1}{c}{\mbox{(T)}}\\
\hline
$1^*$  &0.16           &              &8.85      &-5.94     &\\
\hline
$2$    &0.22           &0.22          &8.70      &-5.85     &3.8\\
$2^*$  &0.22           &              &          &          &\\
\hline
$3^*$  &0.27           &              &8.85      &-5.94     &\\
\hline
$4$  &0.25           &0.30          &9.10      &-6.11     &4.3\\
$4^*$    &0.28           &              &8.90      &-5.99     &4.5\\
\hline
$5$    &0.48           &0.50          &9.65      &-6.48     &4.4\\
$5^*$  &0.50           &              &9.65      &-6.48     &4.4
\end{tabular}
\end{ruledtabular}
\end{table}

\subsection{Magnetization in dc magnetic fields\label{subsec:dc}}
Magnetization data in slowly varying magnetic fields (so-called ``dc fields'')
were taken with a vibrating sample magnetometer (VSM).
The VSM operated in $18\unit{T}$ superconducting magnets.
The sample was in direct contact with a liquid ${}^3\mathrm{He}$ bath, which
was in an insert dewar.
The temperature $0.55\unit{K}$ was reached by pumping on the ${}^3\mathrm{He}$
bath.
The field-sweep time (zero to $18\unit{T}$) was about $1\unit{hour}$.

\subsection{Differential Susceptibility in pulsed magnetic
fields\label{subsec:pulse}}
The differential susceptibility, $dM/dB$, was measured in pulsed magnetic
fields up to $50\unit{T}$ ($500\unit{kG}$).
The techniques have been described earlier.\cite{Shapira99}
The shape of the field pulse ($B$ versus time) was approximately a half cycle
of a weakly damped sine wave, with a rise time  of $3.1\unit{ms}$, and fall
time of $4.3\unit{ms}$.
For each sample, data were first taken with the sample in the pickup coils,
and shortly thereafter with the sample outside the pickup coils.
The signal from the sample was obtained by taking the difference.

The powder samples used in the pulse field experiments were obtained by
crushing the mm-size needles of the growth products.
Each sample, consisting of $20$ to $30\unit{mg}$ fine powder (grain size less
than $0.1\unit{mm}$),  was placed in a thin-walled ($0.25\unit{mm}$)
cylindrical capsule made of Delrin.
The capsule was immersed in a liquid ${}^4\mathrm{He}$ bath which was
maintained at $1.5\unit{K}$.
The Delrin capsule had a small hole at its bottom.
The hole (covered by a tissue paper) allowed a direct contact between the
sample and the bath of superfluid helium.
However, previous experiments have indicated that  despite such a direct
contact the sample may not be in  thermal equilibrium with the bath during
the $7.4\unit{ms}$ field pulse.\cite{Shapira99,Shapira01}

\section{MAGNETIZATION IN DC MAGNETIC FIELDS\label{sec:dcmag}}
\subsection{Experimental Results\label{subsec:expdc}}
\subsubsection{Gross Features\label{subsubsec:grossdc}}
\begin{figure}[tb]\begin{center}
\includegraphics[width=75mm, keepaspectratio=true, clip]{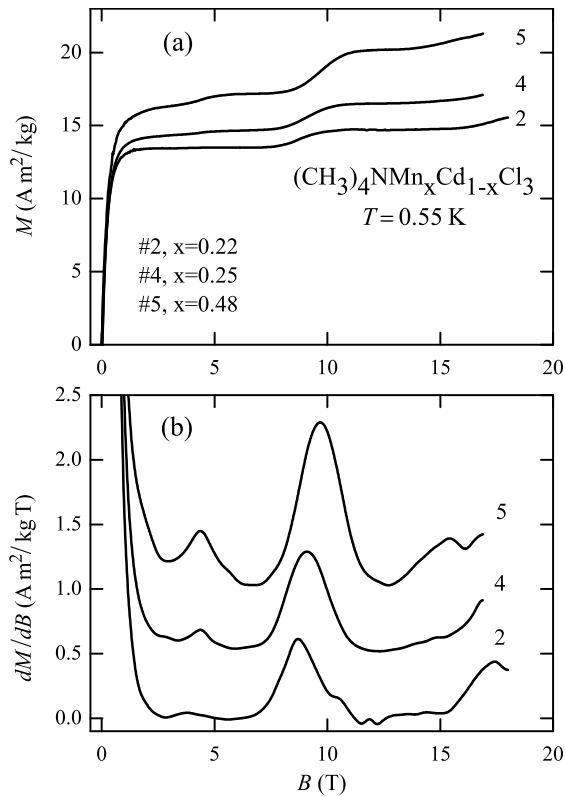}\\
\end{center}\caption{\label{fig6}
(a) Magnetization $M$ of samples  2 ($\x{=}0.22$), 4 ($\x{=}0.25$), and 5
($\x{=} 0.48$), measured at $0.55\unit{K}$ in dc magnetic fields.
The results have been corrected for lattice diamagnetism and addenda.
The SI unit $\mathrm{Am^2/kg}$ is equivalent to $1\unit{emu/g}$.
(b) The numerical derivative, $dM/dB$ of the magnetization traces.
The derivative curves for samples 4 and 5 have been shifted upwards.}
\end{figure}

Figure~\ref{fig6}(a) shows magnetization data at $0.55\unit{K}$ for samples
2, 4, and 5.
These ``dc data'' were actually obtained with a sweep rate of
$\sim0.3\unit{T/min}$.
No hysteresis was observed.
The (very small) corrections for lattice diamagnetism and addenda are included
in Fig.~\ref{fig6}, so that the magnetization $M$ is that of the
$\mathrm{Mn^{2+}}$ ions.
The numerical derivatives, $dM/dB$, of these curves are shown in
Fig.~\ref{fig6}(b).
The Mn concentrations $\x$, measured on the very same samples, are given in
Table~\ref{table1}.

The main features in Fig.~\ref{fig6}(a) are: 1) a fast rise of $M$ at low H,
2) a large MST near $9$ or $10\unit{T}$, and 3) a smaller MST near $4\unit{T}$.
There are also
indications of other MST's at higher fields. For example, sample 2, for which
the data extend to slightly higher fields than for the other samples, shows the
beginning of a large MST near the top of the field. These main features
of the dc magnetization curves  agree with theoretical predictions
for the equilibrium magnetization, such as those in Fig.~\ref{fig3}(b).
The range of the
reduced field $b$ that corresponds to the experimental data in Fig.~\ref{fig6}
extends up to about $4$. The exact maximum value of $b$ is slightly different for
different curves in this figure.

The fast rise of the $M$ at low $B$ corresponds to the alignment of the
zero-field-ground-state spin, $S_T(0)$, of finite chains  with odd $n$. The main
contribution to this fast rise is from the singles ($n{=}1$). The large MST
observed near $9$ or $10 \unit{T}$ corresponds, essentially, to the first MST
from pairs (clusters with $n{=}2$). Other contributions to this
observed MST are from some
longer chains  that have a MST at nearly the same field. For example,  the
second MST  from  quartets (chains with $n{=}4$) is predicted to occur at a field
which is only $2\%$ higher than that of the first MST from pairs. Neither the
triplets ($n{=}3$) nor the quintets ($n{=}5$) have a MST near this field.
The total contribution
of chains with $n{>}2$ to the observed MST near $9$ or $10\unit{T}$ is expected
to be smaller than the contribution from the pairs. The reason is that for
$\x{=}0.5$ the populations $N_n$ of these longer chains are
small compared to the population $N_2$ of the pairs.

For sample 2, a substantial portion of the second MST from pairs is also
seen at the highest fields.
The peak in the derivative $dM/dB$ for this sample, near the top of the
field in Fig.~\ref{fig6}(b), is close to the expected field for this MST,
i.e., $B_2{=}2B_1{=}17.4\unit{T}$. The beginning of the second MST from pairs
is also seen in the derivative curve for sample 4.

The relatively small MST near $4\unit{T}$, seen in Figs.~\ref{fig6}(a) and (b),
is identified as the first MST from the chains with $n {=} 4$ (so-called
``string quartets'').
The first MST from the quartets  is predicted to occur at
$B{=}0.475B_1$, where $B_1$ is the field at the
first MST from pairs.
The experimental results
in Table~\ref{table1} are in reasonable agreement
with this prediction.
The second  MST from quartets which, as already noted, is predicted to occur at
a field which is only $2\%$ higher than $B_1$,
 was not resolved at $0.55\unit{K}$.  This was expected because at $0.55\unit{K}$ the
broadening of any MST due to the finite temperature (``thermal broadening'') is
more than $10\%$ of $B_1$.  The third MST from the quartets is also predicted to
occur within the field range of Fig.~\ref{fig6}. However,  the predictions for
this MST were not fully confirmed by the data.
For example, in Fig.~\ref{fig6}(b) the derivative curve for sample 5
exhibits a small peak near $15\unit{T}$.
For the same sample, the predicted field at the third MST
from the quartets is higher  by about $1\unit{T}$.

Some features of the experimental results in Fig.~\ref{fig6} depend on the Mn
concentration $\x$. The first MST from the quartets stands out more clearly as
$\x$ increases. This trend is expected from the probability curves in Fig.~\ref{fig1}.
The cluster populations $N_n$ are related to these probabilities by Eq.~(\ref{eq:two}).
As $\x$ increases there is an increase in population ratios $N_4$/$N_2$ between
quartets and pairs, and $N_4$/$N_1$ between quartets and singles. Therefore, as
$\x$ increases, the MST from quartets stands out more clearly in comparison with
the MST from pairs, and also in comparison with fast magnetization rise at low
fields.

Another feature that depends on $\x$ is the value of the magnetic field at the
large MST near $9$ or $10\unit{T}$. In Fig.~\ref{fig6}(b) the peak associated with this
MST shifts to slightly higher fields as $\x$ increases. The field at this peak is
expected to be very close to $B_1$. Numerical values as a function of $\x$ are
listed in Table~\ref{table1}. The change of $B_1$ is attributed to a slight dependence of
the NN exchange constant $J$ on the Mn concentration.

\subsubsection{NN Exchange Constant\label{subsubsec:NNJ}}
The NN intrachain exchange constant $J$ was obtained from the
field $B_1$ of the first MST from pairs, using
Eq.~(\ref{eq:fifteen}) and assuming $g{=}2.00$ for the
$\mathrm{Mn^{2+}}$ ion. Values of $J$ for all the samples are
given in Table~\ref{table1}. The exchange constant for $\x{=}0.5$,
$J/\kB  {=} -6.5\unit{K}$, is about $10\%$ higher than for
$\x{=}0.22$. This $10\%$ change is too large to be accounted for
by the unresolved MST from the quartets. As already mentioned,
$J/\kB\cong-6.6\unit{K}$ for pure TMMC ($\x{=}1$). We are not
aware of any theoretical calculation of the $\x$-dependence of $J$
in this system.

\subsection{Comparison with Simulations\label{subsec:sim}}
\begin{figure}[tb]\begin{center}
\includegraphics[width=75mm, keepaspectratio=true, clip]{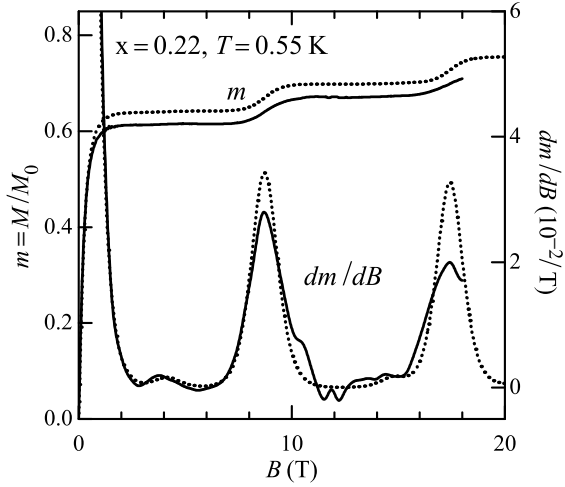}\\
\end{center}\caption{\label{fig7}
Comparison between the dc data for $\x{=}0.22$ at $0.55\unit{K}$ with a
simulation based on equilibrium theory for the single-$J$ model.
The left ordinate scale is for the normalized magnetization $m {=} M/M_0$,
where $M_0$ is the true saturation value.
The right ordinate scale is for $dm/dB$.
Solid curves are from the data in Fig.~\ref{fig6} and the calculated $M_0$.
The dotted curves are from the simulation.
Only thermal broadening, at the actual temperature $0.55\unit{K}$, is included
in the simulation.
}
\end{figure}

\begin{figure}[tb]\begin{center}
\includegraphics[width=75mm, keepaspectratio=true, clip]{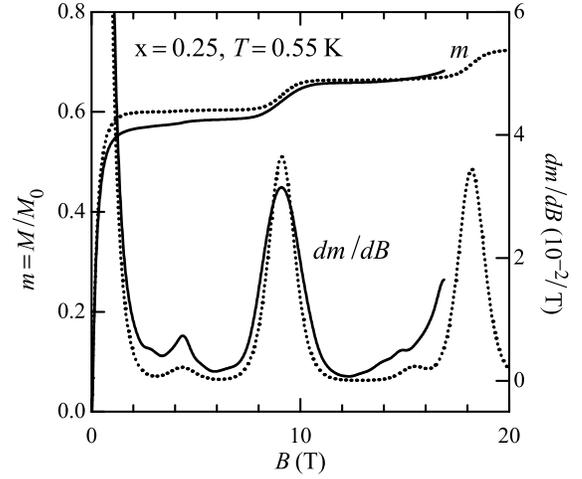}\\
\end{center}\caption{\label{fig8}
Comparison between the dc data for $\x{=}0.25$ at $0.55\unit{K}$
(solid curves) with a simulation based on equilibrium theory for the
single-$J$ model (dotted curves).
The ordinate scales are for $m {=} M/M_0$, and $dm/dB$.
}
\end{figure}

\begin{figure}[tb]\begin{center}
\includegraphics[width=75mm, keepaspectratio=true, clip]{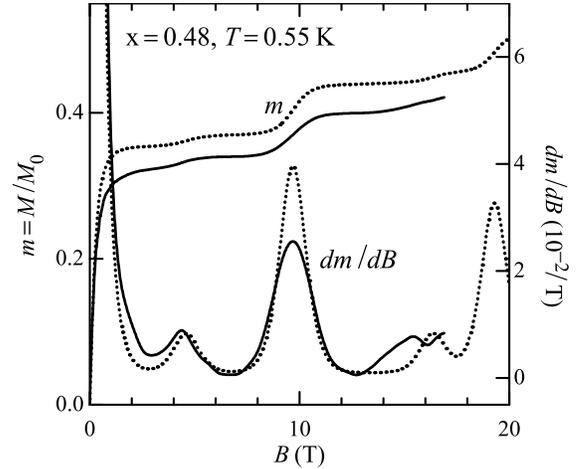}\\
\end{center}\caption{\label{fig9}
Comparison between the dc data for $\x{=}0.48$ at $0.55\unit{K}$
(solid curves) with a simulation based on equilibrium theory for the
single-$J$ model (dotted curves). The ordinate scales are for
$m {=} M/M_0$, and $dm/dB$.
}
\end{figure}

Figures \ref{fig7}--\ref{fig9} compare the experimental results
with numerical simulations based on equilibrium theory for the single-$J$
model (Section~\ref{sec:equilib}).
The comparison is for the normalized magnetization $m {=} M/M_0$, where $M_0$
is the true saturation magnetization.
The ``experimental'' curves use the measured $M$ (Fig.~\ref{fig6}) and the
calculated saturation magnetization $M_0$ for the Mn concentration $\x$.
The simulations assume a random distribution of the Mn ions, and use the
values of $J$ and $\x$(Suscept.) given in Table~\ref{table1} for that sample.
Clusters with $n\le5$ are treated exactly, and the rise-and-ramp approximation
is used for the total contribution of larger clusters.
There are no adjustable parameters in the simulations.

For sample  2, with $\x {=} 0.22$, the agreement between experiment and
theory is reasonably good (Fig.~\ref{fig7}).
The difference between the measured and simulated magnetizations is a few
percent.
It is comparable to the total experimental uncertainty, mainly from the
uncertainty in $\x$. The observed magnitude of the first MST from the pairs is in
agreement with the simulation. However, the associated experimental $dM/dB$ peak
is somewhat broader than in the simulation. Physical mechanisms that broaden
MST's were discussed in Ref.~\onlinecite{Shapira02jap}. In the present case
thermal broadening at the experimental temperature, $T{=}0.55\unit{K}$, is expected to
be the strongest of these mechanisms. It is the only broadening mechanism that
was included in the simulations. Non-thermal causes of line broadening include
the dipole-dipole
interaction and local strains associated with the random replacement of Mn by
Cd. Because  non-thermal broadening  was neglected in the simulation, it is not
surprising that the experimental $dM/dB$ peaks are somewhat broader.
The numerical differentiation of $M$  with respect to $B$ also broadens the
experimental
peak slightly.

Figure \ref{fig8} shows that for $\x {=} 0.25$ the agreement between
experiment and simulation is, again, reasonably good.
However, the observed $dM/dB$ peak at $4.3\unit{T}$, from the first MST for
quartets, is somewhat larger than expected.
The simplest interpretation is that the number of quartets is larger than
given by a random distribution.  The behavior of the derivative $dm/dB$ near
$15\unit{T}$ is attributed to the third MST from quartets.

Figure~\ref{fig9} shows that the agreement between experiment and
simulation for $\x{=}0.48$ is only fair.
The measured magnetization is somewhat smaller than expected, over the entire
field range.
In particular, the initial rise of the measured magnetization is smaller than
in the simulation.
This discrepancy suggests that the Mn ions have a slight tendency to bunch
together, compared to a random distribution.\cite{Shapira02jap}
Also, the peak of $dm/dB$ near $15\unit{T}$  is at a lower field than
predicted for the third MST from the quartets.
The reason for this behavior is unclear. In our view, despite these
discrepancies the agreement between experiment and theory is still reasonable.

\section{DIFFERENTIAL SUSCEPTIBILITY IN PULSED FIELDS\label{pulse}}
Pulsed-field data for $dM/dB$ were taken on powder samples with
$\x\thickapprox0.50$ (from the same product as samples  $5$ and $5^*$),
$\x\thickapprox0.22$ (same product as  $2$ and $2^*$), and $\x\thickapprox0.16$
(same product as  $1^*$).
As noted in Sec.~\ref{sec:experimental}, each sample was in direct contact
with a liquid-helium bath, maintained at $T _\mathrm{bath} {=}  1.5\unit{K}$.
However, such a direct contact  does not ensure
thermal equilibrium with the bath during the $7.4\unit{ms}$ pulse.

The pulsed-field data are presented next.
Non-equilibrium features are pointed out, and many of them are then
interpreted.
However, some aspects of the non-equilibrium behavior are still not fully
understood.

\subsection{Experimental Results in Pulsed Fields\label{exppulse}}
Figure \ref{fig10} shows some $dM/dB$ data for $\x\thickapprox0.50$.
Part (a) of this figure shows an ``up trace'' (increasing $B$) and a
``down trace''
(decreasing $B$) obtained during the same field pulse. Part (b) gives an
expanded view of a portion of the down trace.

Results for $\x\thickapprox0.22$ are shown in Fig.~\ref{fig11}(a).
This figure covers the wide range of $dM/dB$ values that is required to
display the large hysteresis in low fields.
Data taken during the same field pulse, but which cover a much narrower range
of $dM/dB$ values, are shown in Fig.~\ref{fig11}(b).

Results obtained for $\x\thickapprox0.16$ during a pulse with a maximum
magnetic field $B_\mathrm{max} {=}17\unit{T}$ are shown in
Fig.~\ref{fig12}(a).
Figure~\ref{fig12}(b) shows the decreasing-field portion of a trace obtained
during another field pulse for which the maximum field was $34\unit{T}$.
[Unlike all the other pulsed-field data in the present paper, the data in
Fig.~\ref{fig12}(b) have not been corrected for the monotonic background,
because no ``background shot'' (with the sample out of the pickup coils) was
taken in this case.]
Figure~\ref{fig13} shows an expanded, and slightly smoother, view of the
field-down portion of the Fig.~\ref{fig12}(a).

\subsection{\label{subsec:Discussion}Discussion of non-equilibrium effects}
The pulsed-field data do not provide any new information about the exchange
constant $J$.
However, these data show interesting non-equilibrium effects.
The discussion below focuses primarily on:
1) those features of the data that indicate the absence of thermal equilibrium;
2) the change of the non-equilibrium behavior with the Mn concentration $\x$; and
3) physical mechanisms that can give rise to such non-equilibrium effects.

\subsubsection{$\x\thickapprox0.50$}

\begin{figure}[tb]
\begin{center}
\includegraphics[width=75mm, keepaspectratio=true,clip]{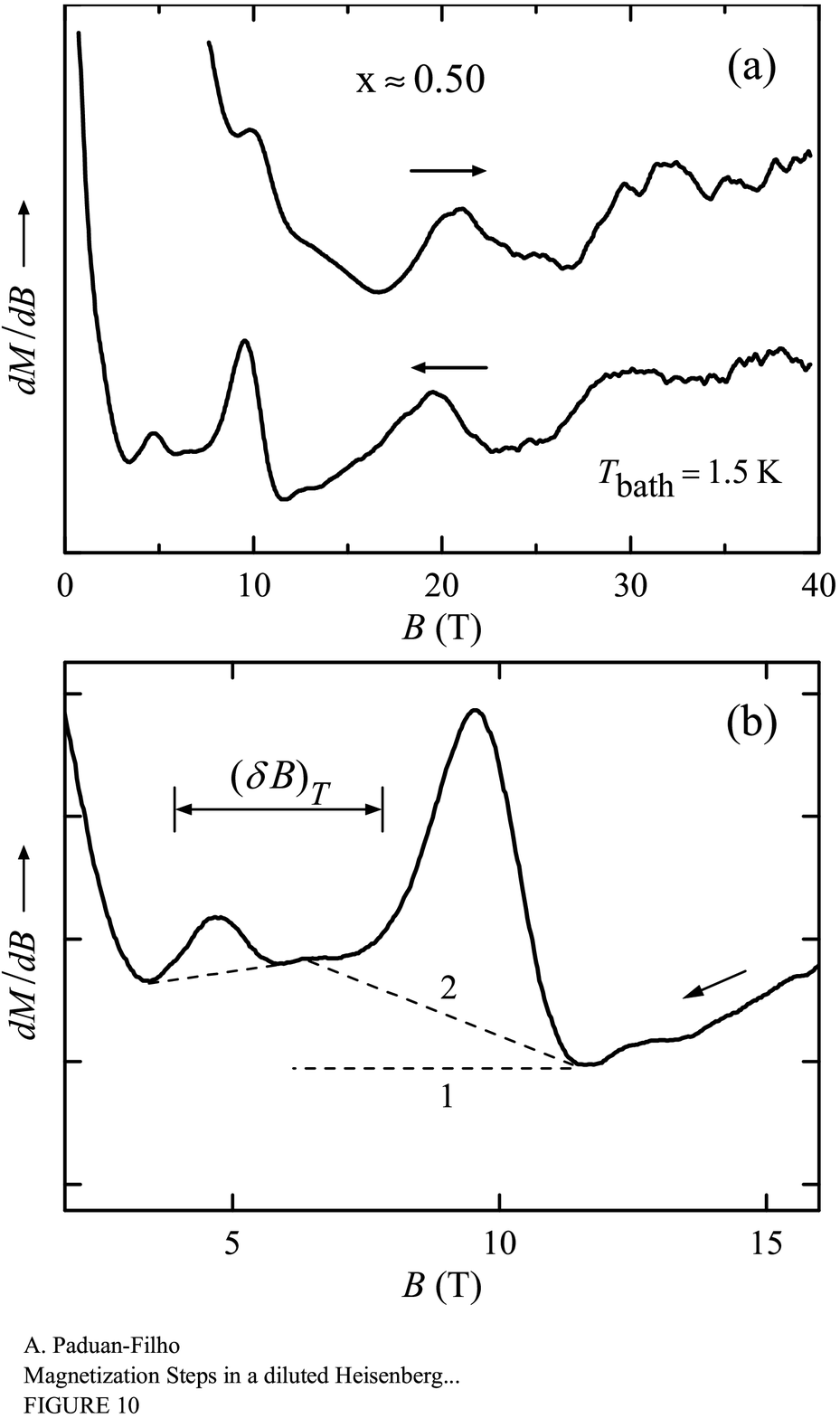}\\
\end{center}
\caption{\label{fig10} (a) Pulsed-field data of $dM/dB$ for
$\x\thickapprox0.50$ in both increasing $B$ (``up'') and
decreasing $B$ (``down''). The up and down traces are shifted
vertically relative to each other. (b) Expanded view of a portion
of the down trace. The calculated thermal width $(\delta B)_T$ at the bath
temperature $T_\mathrm{bath}{=}1.5\unit{K}$ is indicated.
Dashed lines show some choices of baselines used to obtain the experimental widths of two peaks.
}
\end{figure}

Consider first those pulsed-field results in Fig.~\ref{fig10} that are below
$15\unit{T}$. There is a prominent peak near $10\unit{T}$. It stands out more
clearly in the down trace (decreasing $B$) than in the up trace. This peak
corresponds to the first MST from pairs. The down trace also shows a
smaller peak just below $5\unit{T}$. This smaller peak corresponds to the first
MST from the quartets. Although both of these peaks were also observed in the dc
data for the equilibrium magnetization
[Fig.~\ref{fig6}(b)], at least two features of the pulsed-field data indicate
departures from equilibrium with the helium bath, at $T_\mathrm{bath}{=}1.5\unit{K}$.
First, the up and down traces below $15\unit{T}$ are different, i.e.,
there is an hysteresis in this field range.
Second, the widths of both the peak near $10\unit{T}$ and near
$5\unit{T}$ are substantially smaller than the equilibrium width at the bath
temperature.

The ``width'' of a MST will always refer to the full width at half maximum of the
associated peak in $dM/dB$. The various contributions to the equilibrium width
were discussed in Ref.~\onlinecite{Shapira02jap}.
Often, temperature broadening is important.
The thermal width at the temperature $T$ is
\begin{equation}
                (\delta B)_T = 3.53\kB T/g\muB.\label{eqdeltaB}
\end{equation}
Because non-thermal broadening mechanisms are also present in
equilibrium, the thermal width is a lower limit for the actual equilibrium
width. The calculated thermal width, $(\delta B)_T {=} 3.9\unit{T}$ for the actual bath
temperature $T_\mathrm{bath}{=}1.5\unit{K}$, is shown in Fig.~\ref{fig10}(b).

To obtain the experimental width it is necessary to choose a baseline for the
peak in $dM/dB$. Such a choice is not always obvious. In Fig.~\ref{fig10}(b),
two possible choices for the peak near $10\unit{T}$ are indicated by the dashed
lines 1 and 2. Baseline 1 leads to a full width at half maximum of
$2.2\unit{T}$. Baseline 2 leads to a width of $1.9\unit{T}$. On this basis we
conclude that the experimental width is substantially smaller than the
equilibrium width.

The choice of a baseline for the peak near $5\unit{T}$ is
also not obvious. The particular choice shown as a dashed line leads to a width
of $1.2\unit{T}$. Although other choices may lead to a larger experimental
width, it seems that any reasonable choice will lead to a width that is smaller
than the thermal width. Thus, the experimental widths of both of the peaks in
Fig.~\ref{fig10}(b) are smaller than the width that would have occurred had the
sample been in equilibrium with the helium bath.

Figure~\ref{fig10} also shows the second MST from pairs, near
$20\unit{T}$. This  peak is broader than the first peak from the
pairs. A significant feature of this peak is that it  is
asymmetric. As a function of time, the rise of $dM/dB$ as the peak
is approached is faster than the fall after passing through the
peak. This asymmetry is observed in both increasing and decreasing
$B$. An asymmetry of this type  is  expected from models that
assume an inadequate heat flow  (Sec. \ref{subsec:neqmodels}). The
data in Fig.~\ref{fig10} therefore suggest that the
non-equilibrium behavior for $\x\thickapprox0.50$ is due to
inadequate sample-to-bath heat flow.

The IHF scenario can also account for the hysteresis below about
$15\unit{T}$. When the thermal contact with the bath is poor,
strong magneto-caloric effects are expected from the singles
(clusters with $n{=}1$). These effects are analogous to those
involved in the cooling of a paramagnet by adiabatic
demagnetization (and warming by adiabatic magnetization). Of
course, the actual processes in the present case are not truly
adiabatic, because there is some heat flow between the sample and
the bath. In our view, no feature of the data in Fig.~\ref{fig10}
requires CR for its explanation. The observed non-equilibrium
effects seem to be explainable by an inadequate sample-to-bath
heat flow.

Another issue (not directly related to the non-equilibrium behavior)
involves  the third MST from pairs, seen in Fig.~\ref{fig10} near
$30\unit{T}$.
Compared to the first and second MST's from pairs, the third MST is less well
defined.
Specifically, $dM/dB$ hardly decreases on the high-field side of the third MST.
This behavior is explained by the three small MST's from quartets, triplets,
and quintets that are expected between the third and fourth MST's from pairs
(see the MST's near $b {=} 7$ in Fig.~\ref{fig3}).
For $\x{=} 0.5$ the predicted combined size of these three small MST's is
comparable to the size of one MST from pairs.

\begin{figure}[tb]\begin{center}
\includegraphics[width=75mm, keepaspectratio=true, clip]{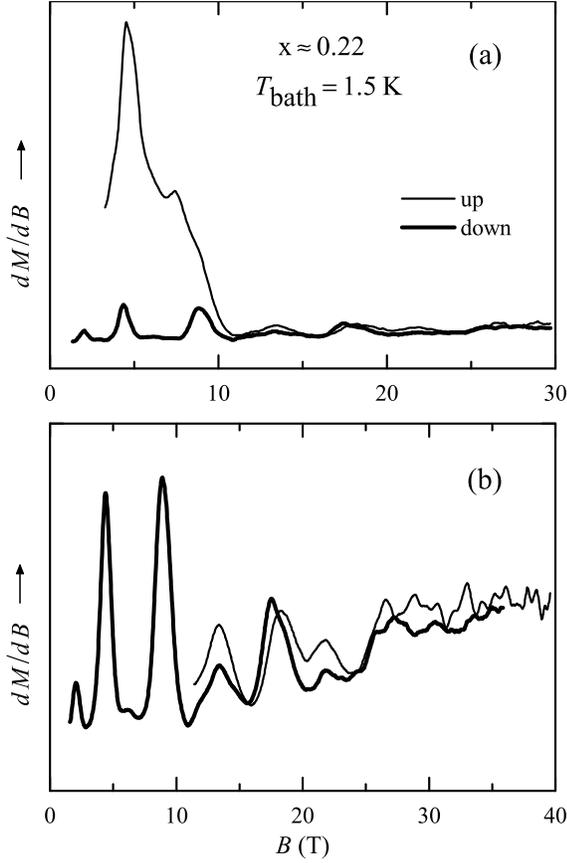}\\
\end{center}\caption{\label{fig11}
Pulsed-field data of $dM/dB$ for $\x\thickapprox0.22$.
(a) Overall view, showing the large hysteresis below about $10\unit{T}$.
(b) Expanded view of the field-up and field-down traces obtained during the
same field-pulse.
The low-field portion of the up trace is excluded.
The up and down traces are not shifted vertically relative to each other.
}
\end{figure}

\subsubsection{Non-equilibrium behavior for $\x\thickapprox0.16$}
\begin{figure}[tb]\begin{center}
\includegraphics[width=75mm, keepaspectratio=true, clip]{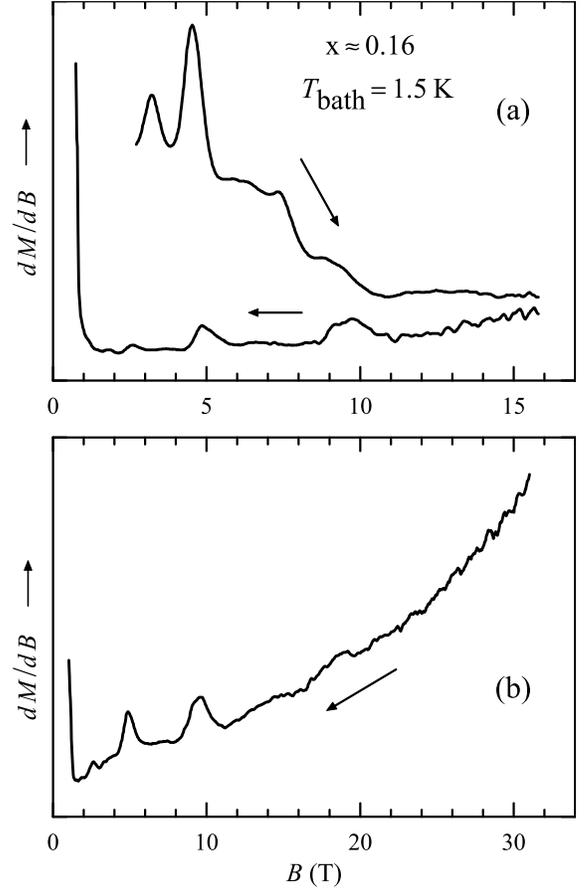}\\
\end{center}\caption{\label{fig12}
(a) Results for $\x\thickapprox0.16$, obtained during a field pulse with a
maximum field $B_\mathrm{max} {=}17\unit{T}$.
The up and down traces are not shifted vertically relative to each other.
(b) The down portion of a trace for $\x\thickapprox0.16$, obtained in another
pulse with $B_\mathrm{max} {=} 34\unit{T}$.
This particular trace, unlike all others, is not corrected for background.
}
\end{figure}

\begin{figure}[tb]\begin{center}
\includegraphics[width=75mm, keepaspectratio=true, clip]{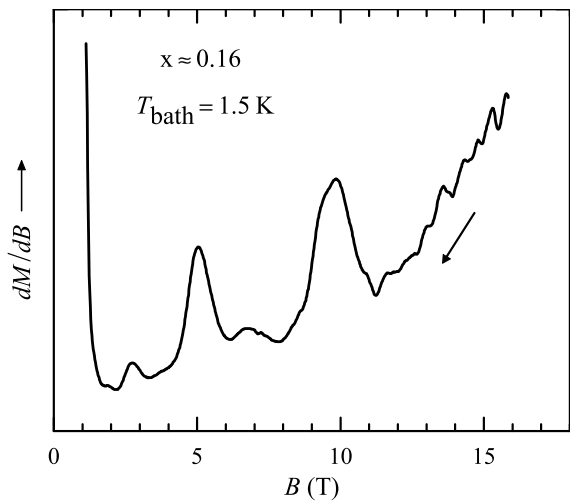}\\
\end{center}\caption{\label{fig13}
An expanded, and slightly smoother, view of the field-down portion of
Fig.~\ref{fig12}(a).
}
\end{figure}

\begin{figure}[tb]\begin{center}
\includegraphics[width=75mm, keepaspectratio=true, clip]{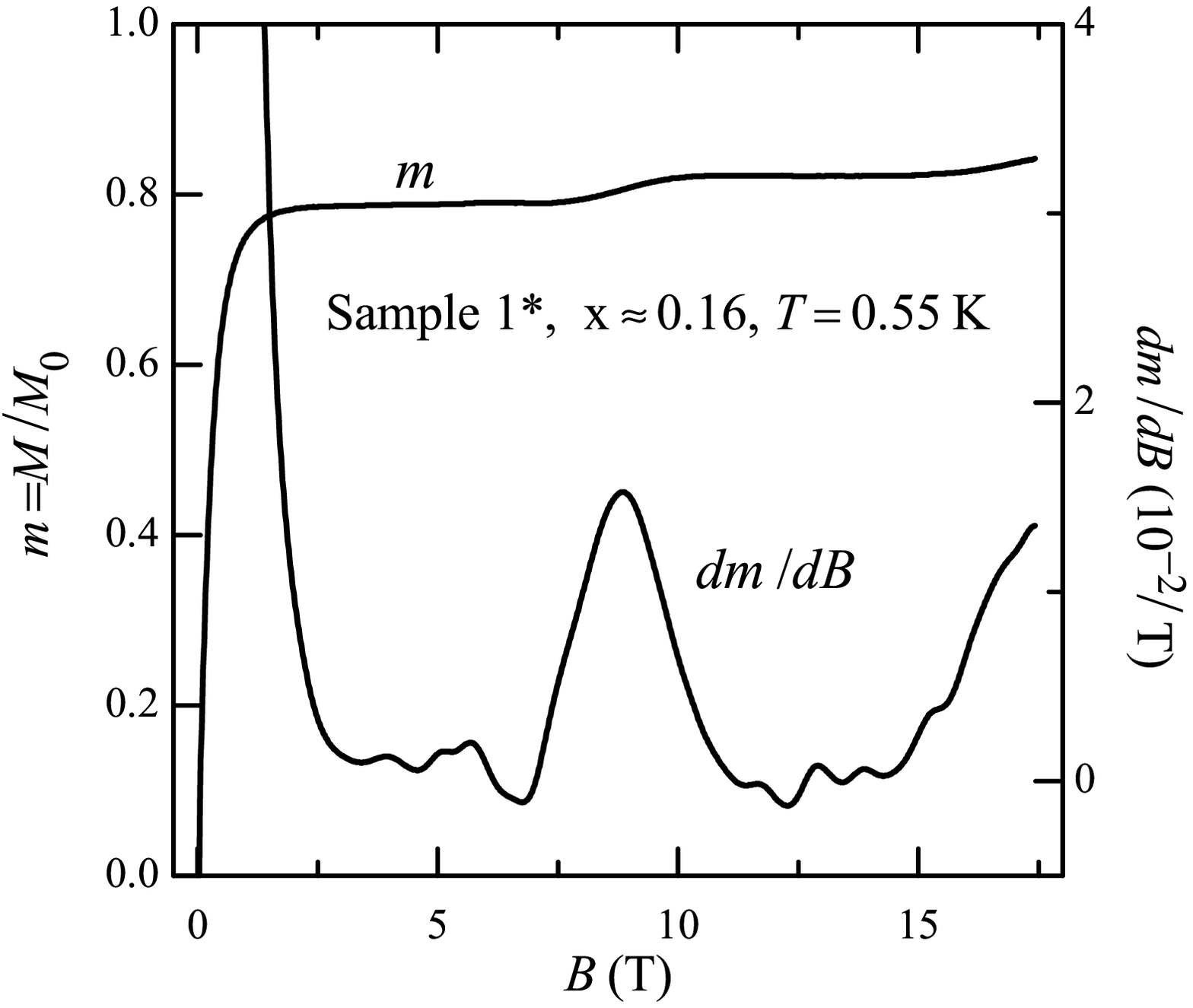}\\
\end{center}\caption{\label{fig14}
The dc magnetization $M$ at $0.55\unit{K}$ for $\x\thickapprox0.16$
(sample $1^*$).
Also shown is the numerical derivative $dm/dB$. In the text, some features of
these equilibrium-magnetization data are contrasted with pulsed-field data on a
similar sample (Figs.~\ref{fig12} and \ref{fig13}).
}
\end{figure}

The absence of thermal equilibrium with the helium bath is very evident in
the pulsed-field data for $\x\thickapprox0.16$.
\begin{enumerate}[(i)]
\item
A pronounced hysteresis is seen in Fig.~\ref{fig12}(a).
\item
In the down portion of the pulse (see Fig.~\ref{fig13}) the widths of the peaks near
$10$, $5$, and
$2.7\unit{T}$ are $1.4$, $0.9$, and $0.6\unit{T}$, respectively.
These values are small compared to a thermal width of $3.9\unit{T}$ at
$T_\mathrm{bath} {=} 1.5\unit{K}$.
A width that is smaller than the thermal width implies a non-equilibrium
behavior.
\item
The peak near $5\unit{T}$ is very pronounced in the pulsed-field data shown in
Figs.~\ref{fig12}(a) and \ref{fig13}, but is barely seen in dc data on a
sample from the same product.
The dc data,  shown in Fig.~\ref{fig14}, should be representative of
equilibrium behavior.
Therefore, the pronounced peak in the pulsed field data is regarded as a
non-equilibrium effect.
\item
The second MST from pairs, near $18\unit{T}$, is barely observed in
Fig.~\ref{fig12}(b), and the third MST from pairs, near $27\unit{T}$, is
totally absent.
The corresponding field-up trace (not shown) exhibits a similar behavior.
Once again the pulsed field data are contrasted with the  equilibrium
magnetization data in Fig.~\ref{fig14}. The latter data, which extend up to
$17.5\unit{T}$,  show a significant portion of the second MST from pairs,
and they indicate that in equilibrium the sizes of the second and first MST's
from pairs are comparable. That is, unlike the behavior in pulsed fields,
the second MST is not small compared to the first.
In equilibrium, all MST's from pairs are expected to be comparable, which is
inconsistent with the results of Fig.~\ref{fig12}(b).

\end{enumerate}

\subsubsection{IHF and CR Scenarios for $\x\thickapprox0.16$}
The large hysteresis at low fields [Fig.~\ref{fig12}(a)] is not uncommon in
pulsed field experiments on diluted magnetic materials.\cite{Foner94,Ajiro98}
In the present case the hysteresis is largely due to the slow
response  of  the singles to the rapidly changing $B$.
In the up portion of the pulse the alignment of the  spins of the singles is not
completed until $B$ is above $10\unit{T}$.
In the down portion of the pulse these spins remain largely aligned until $B$
drops below $1\unit{T}$.
For $\x\thickapprox0.16$, the majority of the spins ($70\%$ for a random
distribution) are singles,  so  that the hysteresis is very pronounced.

MST's from pairs and larger clusters stand out more clearly in the down
portion of the field pulse because the singles remain largely aligned until
$B$ drops below $1\unit{T}$.
For this reason the field-down portion of the traces in Figs.~\ref{fig13}
and \ref{fig12}(b) is discussed first.

Spin-lattice relaxation times are often longer for lower $\x$, which is the
likely cause of the more pronounced non-equilibrium behavior for
$\x\thickapprox0.16$ compared to $\x\thickapprox0.50$.
An important issue in the data interpretation is whether the spin-lattice
relaxation is fast enough to maintain equilibrium {\em within} the sample.
In that case the IHF scenario would apply (see Sec.~\ref{subsubsec:IHF}).
An alternative is a more severe non-equilibrium behavior which is better
described by  the CR scenario, including single spin flips near level
crossings of excited states (Sec.~\ref{subsubsec:cross}).%
\cite{Ajiro98,Wernsdorfer02}

Below we consider both the IHF and CR scenarios for
the down portion of the pulse. The preferred scenario cannot be chosen
on the basis of the fields at which the MST's occur, because the differences
are often small compared to the experimental accuracy.
However, the two scenarios lead to different relative sizes of the peaks in
$dM/dB$. The poor agreement of the observed relative sizes with those predicted
by the IHF scenario will suggest that the CR scenario is preferable for this Mn
concentration.

Consider first the IHF scenario. In this  scenario the magnetization of each
cluster type is the equilibrium magnetization at $T_s$.
For the low $T_s$ indicated by the small widths of the observed MST's (in the
down portion of the pulse), this magnetization is that of the ground state.
The largest peak in Fig.~\ref{fig12}, just below $10\unit{T}$, is mainly due
to the first MST from pairs.
The second largest peak, near $5\unit{T}$ is the first MST from quartets.
It is predicted to occur at a field which is $0.48$ times that of the first
peak for pairs.
The second $dM/dB$ peak from quartets should be at a field which is $2\%$
higher than the first peak from pairs.
The structure of the peak near $10\unit{T}$ is possibly due to a superposition
of these two MST's, although it is much wider than $2\%$.

In the IHF scenario the field at the first MST for finite chains with even $n$ can be estimated
from the energy levels given in Ref.~\onlinecite{Lou02}.
For $n$ between $4$ and $10$, this field is nearly proportional to $1/n$.
A similar approximate dependence on $n$ holds for AF rings (closed chains).%
\cite{Gatteschi00}
On this basis the small sharp peak near $2.7\unit{T}$ (in Fig.~\ref{fig12})
is due to octets or sextets.
There is also an indication of a small peak near $2.0\unit{T}$, which would be
attributed to larger clusters.

Although the IHF scenario accounts for the fields of many of the observed
MST's, this scenario is very questionable for this Mn concentration.
Assuming that the Mn cations are randomly distributed, the number of quartets
for $\x = 0.16$ is smaller than the number of pairs by a factor of $39$.
For sextets and octets the factors are $1.5\times10^3$ and $6\times10^4$,
respectively.
Therefore, unless the deviations from random distribution are extremely
large, it should not have been possible to observe MST's from sextets or octets
if the behavior followed the IHF scenario.
The first MST from quartets might have been detectable, but it should  have been very
small compared to the MST from pairs.  This was not the case
in the pulse field experiments (Figs.~\ref{fig12} and \ref{fig13}).
As predicted, the equilibrium data in Fig.~\ref{fig14}, for nearly the same $\x$,
indicate that the first MST from quartets is much less pronounced
than the first MST from pairs.

In Fig.~\ref{fig12}(b) the first MST from pairs stands out clearly but the
second MST from pairs is barely visible, and
the third is totally absent.
These results also are not well understood within the IHF scenario, although
some indication of such a behavior appeared in simulations by Nakano and
Miyashita for iron clusters with ring structure.\cite{Nakano01}

In the CR scenario the MST's in Figs.~\ref{fig12}(b) and \ref{fig13}
(both for  decreasing B) are interpreted as follows.
The peak just below $10\unit{T}$ is the first fundamental peak $\mathrm{P}_1$,
with some contributions from $\mathrm{P}_{2/2}$, $\mathrm{P}_{3/3}$, etc.
The peak at $5\unit{T}$ is the second harmonic peak $\mathrm{P}_{1/2}$.
The small peak at $2.7\unit{T}$ is the third-harmonic peak $\mathrm{P}_{1/3}$
or the fourth-harmonic peak $\mathrm{P}_{1/4}$.
The broad peak near $6.6\unit{T}$ is possibly $\mathrm{P}_{2/3}$.
Because this interpretation uses only pairs and singles,\cite{Ajiro98} it is
not open to objections  based on the low populations of quartets and larger
clusters.
The CR scenario also accounts for some features of the up trace in
Fig.~\ref{fig12}(a).
The large peak near $5\unit{T}$ is attributed to the CR process of the type
shown in Fig.~\ref{fig4}(b), except that the directions of both spin flips
are reversed.
This process allows the singles to relax toward a state with a higher
magnetization.
The magnetization of the pairs also increases by this process.

The process in Fig.~\ref{fig4}(b) involves only one single and one pair,
and is the simplest CR process between singles and pairs.
It accounts for the $\mathrm{P}_{1/2}$.
More complicated CR processes can lead to other ``harmonic peaks,'' such as
$\mathrm{P}_{1/3}$, $\mathrm{P}_{2/3}$, $\mathrm{P}_{3/4}$.
These peaks are expected to be smaller than $\mathrm{P}_{1/2}$.
The peak $\mathrm{P}_{1/3}$ may involve processes such as a CR between a pair
and two singles, or between a pair and a single which undergoes a double spin
flip.
The peak $\mathrm{P}_{2/3}$ may involve a spin flip in a single and spin flips
in two pairs, etc. In the up trace the observed peak at $3.2\unit{T}$, and the
small peaks at $6.4$, and $7.4\unit{T}$ may correspond to the
$\mathrm{P}_{1/3}$, $\mathrm{P}_{2/3}$ and $\mathrm{P}_{3/4}$ harmonics.
The peak observed near $10\unit{T}$ is a superposition of  the first
fundamental peak from pairs, $\mathrm{P}_{1}$, and the harmonics
$\mathrm{P}_{2/2}$, $\mathrm{P}_{3/3}$, etc.

As already noted, in Fig.~\ref{fig12}(b) the second MST from pairs is
barely visible, and the third is totally absent.
These results suggest that for $\x\thickapprox0.16$, the spin relaxation for
pairs in fields above $15\unit{T}$ is  very slow compared to a millisecond.
As $B$ sweeps through a region where a MST from pairs should have occurred,
the pairs are unable to relax towards the new ground state.
We speculate that the slow spin relaxation for pairs is mainly due to a
reduction of CR between pairs and singles, and that this reduction is related
to the saturation of the singles in fields above $15\unit{T}$.
CR between different pairs, or between pairs and larger clusters, is expected
to become slower as $\x$ decreases.\cite{Abragam70}
Among the three samples, such CR processes should be least efficient for
$\x\thickapprox0.16$.

\subsubsection{Non-equilibrium behavior for $\x\thickapprox0.22$}
Non-equilibrium behavior is also evident for $\x\thickapprox0.22$.
The large low-field hysteresis for this Mn concentration, in
Fig.~\ref{fig11}(a), is somewhat similar to the hysteresis in
Fig.~\ref{fig12}(a) for $\x\thickapprox0.16$.
In the up trace, the large peak near $4.5\unit{T}$ is identified as the
$\mathrm{P}_{1/2}$ peak, and is attributed to the cross relaxation process in
Fig.~\ref{fig4}(b), with the arrows reversed.

The down trace in Fig.~\ref{fig11}(b) exhibits large peaks near $8.9\unit{T}$
and $4.4\unit{T}$ approximately, and a small peak near $2.1\unit{T}$.
The widths at half height of these peaks, $1.3$, $0.9$, and $0.7\unit{T}$,
respectively, are all much smaller than thermal width of $3.9\unit{T}$ at
$T_\mathrm{bath}$.
These widths, which are similar to those for $\x\thickapprox0.16$,
indicate non-equilibrium behavior.

The largest peak in Fig.~\ref{fig11}(b), at $8.9\unit{T}$, is undoubtedly
the first fundamental peak from pairs, $\mathrm{P}_1$.
The second fundamental peak $\mathrm{P}_2$ is also observed near $18\unit{T}$.
The CR scenario predicts large second-harmonic peaks $\mathrm{P}_{1/2}$,
$\mathrm{P}_{3/2}$, $\mathrm{P}_{5/2}$ at $4.5$, $13.4$, and $22.3\unit{T}$,
respectively.
These fields are close to $4.4$, $13.4$, and $21.8\unit{T}$, of large peaks in
Fig.~\ref{fig11}(b).
The peak observed near $2.1\unit{T}$ is consistent with $\mathrm{P}_{1/4}$.
It is possible, but far from certain, that the small peaks at $6.2$ and
$12.0\unit{T}$ are $\mathrm{P}_{2/3}$ and $\mathrm{P}_{4/3}$.

The main difference between  $\x\thickapprox0.22$ and $\x\thickapprox0.16$ is
that for the higher Mn concentration there are still prominent MST's above
$10\unit{T}$.
For $\x\thickapprox0.50$ the second and third MST's from pairs are even more
pronounced [see Fig. \ref{fig10}(a)].
These results suggest that at these fields the spin relaxation rate for pairs
increases rapidly with Mn concentration.
We tentatively attribute this trend to the expected increase with $\x$ of the
efficacy of CR processes involving pairs, and pairs and larger clusters.

\subsubsection{Summary of the analysis of non-equilibrium behavior}
A definitive interpretation of the observed non-equilibrium behavior in pulsed
fields is still lacking.
However, it appears that for $\x\thickapprox0.50$ the non-equilibrium behavior
is better explained by the IHF scenario.
For $\x\thickapprox0.16$ and $\x\thickapprox0.22$ the data are better
explained by spin flips associated with CR and with level crossings.
The data suggest that at the high magnetic fields where the magnetization of
singles is saturated, the spin relaxation rate for pairs increases rapidly
with increasing $\x$.
This increase is tentatively attributed to the $\x$-dependence of CR processes
involving pairs.

\begin{acknowledgments}
We thank R. Muccillo (USP) and J. A. Adario (MIT) for X-rays measurements.
The work in Brazil was supported by FAPESP (Funda\c{c}{\~a}o de Amparo {\`a}
Pesquisa do Estado de S{\~a}o Paulo, Brazil) under contract number
99/10359--7. The authors APF, NFOJ and VB acknowledge support from CNPq
(Conselho Nacional de Desenvolvimento Cient{\'\i}fico e Tecnol{\'o}gico,
Brazil). Travel funds for YS were provided by FAPESP.
\end{acknowledgments}

\begin{thebibliography}{52}
\bibitem[{not({\natexlab{b}})}]{note21}
\bibinfo{note}{Taking the sum up to $n=\infty$ may seem inconsistent with the
  assumed short chain model. However, for the low x considered here (below
  about 0.5), the contribution to the sum in Eq. (\ref{eq:eight}) from chains
  that are not short is negligible.}
\end{thebibliography}
%
\newcommand{\noopsort}[1]{} \newcommand{\printfirst}[2]{#1}
  \newcommand{\singleletter}[1]{#1} \newcommand{\switchargs}[2]{#2#1}

\end{document}